# Multiplexed PSF engineering for 3D multicolor particle tracking

Nadav Opatovski[1], Yael Shalev Ezra[2], Lucien E. Weiss[2], Boris Ferdman[1], Reut Orange[1] and Yoav Shechtman[1,2,*]

[1]*Russel Berrie Nanotechnology Institute, Technion - Israel Institute of Technology, Haifa, 3200003, Israel*
[2]*Department of biomedical engineering, Technion - Israel Institute of Technology, Haifa, 3200003, Israel*
*\*yoavsh@bm.technion.ac.il*

**Abstract:** Three-dimensional spatiotemporal tracking of microscopic particles in multiple colors is a challenging optical imaging task. Existing approaches require a trade-off between photon-efficiency, field of view, mechanical complexity, spectral specificity and speed. Here, we introduce multiplexed point-spread function engineering that achieves photon efficient, 3D, multicolor particle tracking over a large field of view. This is accomplished by first chromatically splitting the emission path of a microscope to different channels, engineering the point-spread function of each, and then recombining them onto the same region of the camera. We demonstrate our technique for simultaneously tracking five types of emitters in-vitro, as well as co-localization of DNA loci in live yeast cells.

## 1. Introduction

Tracking individual fluorescent probes with nanoscale precision has many biological applications, including characterization of membrane dynamics[1], DNA replication and repair, gene expression and regulation[2,3], RNA structure[4,5], and protein folding[6]. Individual trajectories of fluorescently labelled biomolecules have revealed key insights into their behavior and environment[7,8] and also the interaction between two or more is of key interest[9,10], which often necessitates multicolor imaging.

In high numerical-aperture (NA) optical systems, tracking by direct imaging suffers from a shallow depth of field, outside of which emitters appear blurry and dim, and localization precision is compromised. This fundamental limitation has driven the emergence of various methods that enable 3D tracking. Notable techniques such as light sheet microscopy [11,12] and confocal laser-scanning microscopy (CLSM)[13] have shown to provide advantages in resolution and signal to noise ratio (SNR), however, they require a scanning procedure, which inherently impairs temporal resolution, making them sub-optimal for imaging of fast, dynamic scenes.

Scan-free methods for 3D single particle tracking (SPT) include interference-based imaging, multiplane imaging, feedback-control, and point-spread-function engineering (PSFE). Interference-based methods extract the 3D position of an emitter from an interference pattern, usually generated by collecting light using two objective lenses. This approach is typically limited to a small axial range of several hundreds of nanometers due to the ambiguity associated with interference of visible light[14,15]. In multiplane imaging, light from the sample is split to form multiple images corresponding to different focal planes, effectively providing simultaneous views of emitters from different nominal focal planes. Due to splitting the light to multiple images, either multiple cameras are required[16], or the field of view (FOV) is restricted in order for each depth to fit side-by-side on a single camera sensor[17,18]. Feedback-controlled methods are usually characterized by exceptionally high spatial and temporal resolution, but are typically limited to tracking a single particle at a time, or to a very small FOV[19,20].

In PSFE, a special phase mask in the pupil plane is used to modify the wavefront originating from a point source, such that 3D information is encoded in the shape of the PSF[21–23]. This enables scan-free 3D localization by subsequent computational image analysis. An important aspect relevant to all mentioned methods is that for achieving optimal resolution, the optical system must be highly photon-efficient; in localization microscopy, the Poisson-distributed nature of light dictates that localization precision scales as the inverse square-root of the photon count, at best[24–26].

Of key interest in many particle tracking experiments is measuring the relative motions of two or more types of particles, which can be accomplished by conjugating labels with unique spectra and performing multicolor imaging. Sensitive detectors (cameras) in fluorescence microscopy are typically grayscale, therefore the straightforward way to obtain spectral information is to perform sequential image acquisition, naturally entailing a decrease in temporal resolution. An alternative is to dedicate different regions on the sensor, or different camera sensors, to different spectral channels by using dichroic filters or prisms. This approach is photon-efficient and enables simultaneous multicolor acquisition, but significantly reduces the FOV, or is costly and space-consuming[27] if multiple cameras are used. Channel splitting can also be combined with PSFE to encode 3D information[9,10,28].

Extraction of spectral and depth information while minimizing temporal and spatial penalty has been previously demonstrated[29], where color-dependent PSFs encoded the colors of emitters while maintaining the benefits of PSFE. Phase modulation was implemented by a liquid crystal spatial light modulator (LC-SLM) generating a single, wavelength-dependent phase mask (see also[30,31]) yielding two different PSFs for two specified spectral bands. While this approach proved successful for discriminating two colors, a photon-efficient extension to three or more colors remains a difficult task using commercially available LC-SLMs. This is due to three main reasons – first, the large dynamic phase range required; second, the polarization dependency of the LC-SLM requires ~50% photon-loss; third, the broad emission spectra of most fluorescent molecules is incompatible with the strong wavelength-dependence necessary for a three-color phase mask. An alternative for LC-SLMs are diffractive optical elements (DOEs). Single-color DOEs can be fabricated by photolithography[32], however a wavelength-dependent DOE for multicolor PSFE is challenging to fabricate due to the requirement for a large etching depth range at nanoscale precision. The benefit of encoding spectral characteristics into the PSF optimized for 3D tracking makes it worthwhile to pursue methods to bypass the technical limitations posed by the multicolor LC-SLM or DOE.

Here, we introduce multiplexed PSFE: a method for highly photon-efficient multicolor 3D SPT over a large FOV, by combining multiple color-specific engineered PSFs onto a single imaging channel. The concept of the method is to use simple, affordable DOEs while generating a detectable difference between the PSFs of different colors to provide multicolor detection. Importantly, photon loss is inherently minimal, as the major sources of photon loss of existing large-FOV multicolor PSFE are circumvented, thus video-rate imaging of molecular probes is possible with precision of tens of nanometers. We demonstrate our method by 3D simultaneous tracking of five spectrally distinct populations of diffusing beads in a suspension and DNA loci in live yeast cells.

## 2. The principle of multiplexed PSFE

We achieve multiplexed PSFE for multicolor 3D-SPT by first splitting the image into separate spectral channels using dichroic beamsplitters. Next, PSFE is performed in each channel, and finally the channels are recombined at the camera sensor plane. In traditional PSFE systems, the camera is positioned downstream of a *4-f* extension to the optical path, in order to gain physical access to the Fourier-plane of the optical system. This enables phase modulation of the pupil function by placing a carefully-aligned phase mask in the Fourier plane. In our setup, the extra available space along the extended optical path is used to separate the light into multiple spectral channels, and in each, perform PSFE by a different phase mask. Consequently, both the color and 3D position of a point source are encoded by the shape of the PSF.

In our work, we use the 4 μm-Tetrapod PSF[33], which is designed to provide optimized 3D localization precision over an axial range of 4 μm. The Tetrapod PSF also benefits from appearing significantly different upon rotation of the phase mask by small angles (Fig 1.b). This enables the use of three similar-shaped phase masks at different lateral angles to generate differently rotated, distinguishable Tetrapod PSFs. Throughout the paper, the three channels will be referred to as *blue, orange* and *red*, corresponding the perceived color of their spectral transmission bands upon white-light illumination.

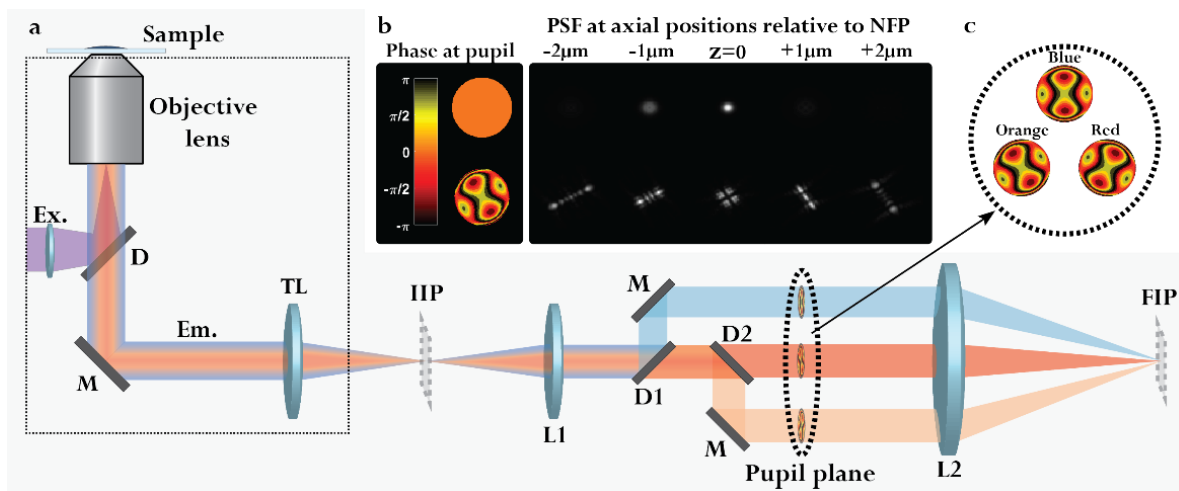

**Fig. 1**: The optical system. (a) A simplified schematic drawing of the optical system. Fluorescence of multiple spectral bands is emitted from the sample forming a colorful image at the intermediate image plane. Subsequently, a 4-f system is placed, inside of which the light is divided to three spectral channels by dichroics. In each channel, PSFE is performed using a channel-specific phase mask. Abbreviations: Ex. - excitation laser, Em. - Emitted fluorescence, D, D1, D2- Dichroic beamsplitters, M - mirror, TL - tube lens, L1, L2 - lenses of the 4-f system, IIP - intermediate image plane, FIP - final image plane (Note: Several mirrors were excluded from the drawing). (b) Concept of PSFE - by introducing a phase mask at the pupil plane, depth is encoded in the shape of the PSF as measured in the final image plane. A simulated comparison between the standard PSF (top row) and the 4 μm Tetrapod PSF (bottom row) is shown. The PSFs are of a sub-diffraction emitter placed at five different axial positions relative to the nominal focal plane (NFP). The orientation of the PSF is directly derived by the orientation of the phase pattern at the pupil, here presented tilted at roughly $\pi/6$. (c) A representation of the phase masks placed at the pupil plane. Each is uniquely oriented to produce a channel-specific PSFs.

## 2.1 Localization

Determining the position of a particle given its measured PSF, i.e. localization, is performed by fitting a model of the PSF to a region-of-interest (ROI). The model is derived from a phase-retrieved pupil function that is obtained by a vectorial implementation of phase retrieval (VIPR)[34]. The parameter fitting is done using maximum likelihood estimation (MLE) assuming Poisson photon statistics.

In the experiments performed here, the fitting procedure is divided into two scenarios based on the background intensity pattern. For imaging fluorescent beads diffusing in a droplet of $H_2O$-glycerol mixture, the background is treated as a constant across the ROI. In this case, the fitted parameters are the 3D position (x,y,z), the total signal photon counts detected from the emitter ($N_{ph}$), and the constant background photon count per pixel (b). When background cannot be assumed constant inside the ROI, namely inside cells, background-shape parameters must be added. In this case, a two-dimensional Gaussian background is added to the model, which is parameterized by the centroid position ($x_g, y_g$), the standard deviations in orthogonal directions ($\sigma_x, \sigma_y$), the angular tilt of the Gaussian ($\theta$), and photon count ($N_{ph-g}$). To reduce the complexity of the fit function, the tilt angle and the constant background term are not fitted - the tilt angle per PSF is defined to be the same as the PSF orientation, due to the Tetrapod PSF modifying the background emission as well, and the constant background is estimated from the ROI periphery.

## 2.2 Image registration

Our imaging system effectively divides the emission light into three individual optical paths where the final images are overlaid on the detector with an intentional small shift. This creates three coordinate systems, and necessitates registration between them. The registration is done by imaging randomly scattered multicolor fluorescent beads (Tetraspeck T7280, Invitrogen), which appear bright in all three channels. The three PSFs of each bead are localized, and a 3$^{rd}$ order, global-polynomial transformation is used to map the coordinate systems onto each other. In order to register the entire imaging volume, the microscope stage is moved in all three dimensions, and multiple images are acquired.

Throughout the experiments, we verified the stability of the registration by recording registration datasets both prior to and after acquisition of the experimental data. In a post processing step, the correspondence between the registrations from both datasets was checked to verify that the transformations generated before the experiment were still valid after the experiment.

## 2.3 Methods

All experiments were performed using an inverted microscope (Ti2 Eclipse, Nikon), equipped with a silicone-immersion objective (SR HP Plan Apo 100X/1.35 Sil λS, Nikon). Images were captured using an sCMOS camera (Prime 95B, Teledyne Photometrics). Excitation lasers were 488 nm and 532 nm for diffusing beads experiments, at optical densities of ~2.5 W/mm² each, and 488 nm and 561 nm for live-cell experiments, at optical densities of ~5.5 W/mm² and ~11.5 W/mm², respectively. The spectral channels of $\lambda < 567$ nm (blue), 567 nm $< \lambda <$ 635 nm (orange) and $\lambda > 635$ nm (red) were set by longpass beamsplitters (Fig. 1) and upon necessity, additional filters were added per channel. In each channel, a photolithography-etched fused silica dielectric phase mask was placed in the Fourier plane (Fig 1.c). The lenses of the *4-f* system were of 200 mm focal length each, providing no additional magnification to the image.

## 3. Results

### 3.1 Five-color 3D tracking

The optical system, although consisting of three PSFE channels, enables identification and tracking of more than three populations using the inter-channel ratiometric information. As proof-of-concept, fluorescent beads of five distinct spectral emission profiles were imaged freely diffusing in an $H_2O$-glycerol mixture and simultaneously imaged at video-rate for tracking (Fig. 2). To our knowledge, 3D tracking of >3 distinct diffusing particle populations over a large FOV at high spatiotemporal resolution was not previously performed. In this experiment, temporal resolution was defined by the camera exposure, at 19 ms/frame (~50 fps), providing high temporal resolution while maintaining sufficient exposure time to provide high localization precision. Additionally, the short exposure times minimized the effect of motion-blur. The properties of the fluorescent beads are summarized in Table 1.

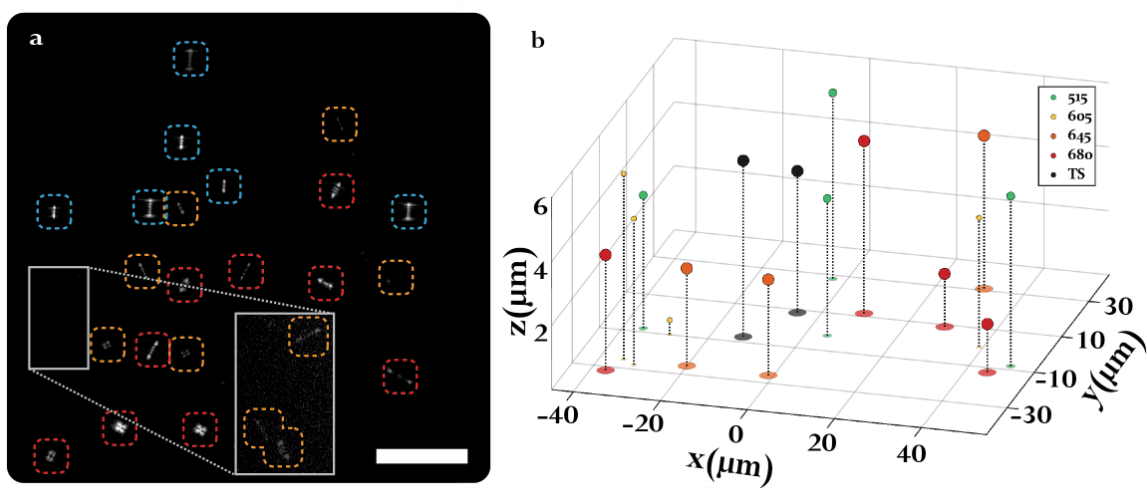

**Fig. 2**: Example of a single frame acquired by the optical system, showing multiple types of freely diffusing beads. (**a**) Experimental image containing multiple PSFs. Each PSF encodes 3D position as well as color (indicated by the dotted box around it). The rectangle shows an improved contrast region containing three dim orange-channel beads. Scale bar = 20 μm. (**b**) 3D positions and colors of the beads from (a). The diameters of the different bead types in the plot are based on manufacturer data (Table 1). The legend states the spectral emission peaks of the beads; TS = Tetraspeck. More detail on bead types is given in Table 1.

In each frame, ROIs suspected as containing a PSF were selected and the PSF type (color) was determined algorithmically based on the orientation of the brightest pixels. Then localization was performed, as explained in the "Localization" section. After each dataset was localized, the PSFs were sorted into trajectories according to color and position, and by finding pairs or triplets of highly correlated trajectories, PSFs of similar beads were recognized. Classification to bead types was based on prior knowledge of the emitters' spectra. This can be calculated theoretically from the spectral curves of the emissions and the filters, or it can be easily experimentally calibrated. In the performed experiment only two beads types were clearly visible in more than one channel: Fluospheres Crimson was apparent in the orange and red channels, and Tetraspeck was clearly apparent in all of the channels (see visualization 1).

In order to quantify the results from the diffusion experiment, per bead type, the mean-square-distance (MSD) of the three shortest time delays was calculated, from which we extracted the diffusion coefficient according to[35]:

$$MSD(\tau) \equiv \frac{1}{N - \frac{\tau}{\Delta t}} \sum_{i=1}^{N-\frac{\tau}{\Delta t}} (\vec{x}(i\Delta t) - \vec{x}(i\Delta t + \tau))^2 = 2nD\tau + c \qquad (1)$$

Where $N$ is the total number of positions measured, $\Delta t$ is the time difference between consecutive measurements and $D$ is the diffusion coefficient, $n=3$ is the dimension of motion, and $c$ a constant that depends on the localization precision and motion blur, both, in fact, functions of $D$. For our analysis $D$ was extracted by fitting a line to MSD as a function of the time-delay $\tau$, based on the three first time-delays ($\tau = \Delta t, 2\Delta t, 3\Delta t$).

Table 1. Bead specs according to manufacturer batch data

| Bead name | Diameter (nm) | Peak emission wavelength (nm) |
|---|---|---|
| FluoSpheres Yellow-green (F8807, Invitrogen) | 99±7.8 | 513 |
| FluoSpheres Red (F10720, Invitrogen) | 46±2 | 607 |
| FluoSpheres Crimson (F8806, Invitrogen) | 210±11 | 644 |
| FluoSpheres Dark red (F8807, Invitrogen) | 220±12 | 678 |
| Tetraspeck (T7280, Invitrogen) | 210±10 | 430, 515, 580, 680 |

As a quantitative measure to validate the tracking procedure, we calculated the MSD of diffusers from the same dataset to extract $D$. According to Stokes-Einstein relation[36] the expected value of $D$ is inversely proportional to $r$, which is known (Table 1). The MSD values, in $\mu m^2/s$, were calculated from the trajectories of a single dataset (Fig. 3) are {0.69±0.28, 1.22±0.39, 0.29±0.09, 0.23±0.10, 0.23±0.08}, according to the order of their appearance in Table 1, showing good correspondence to the theoretical expectations.

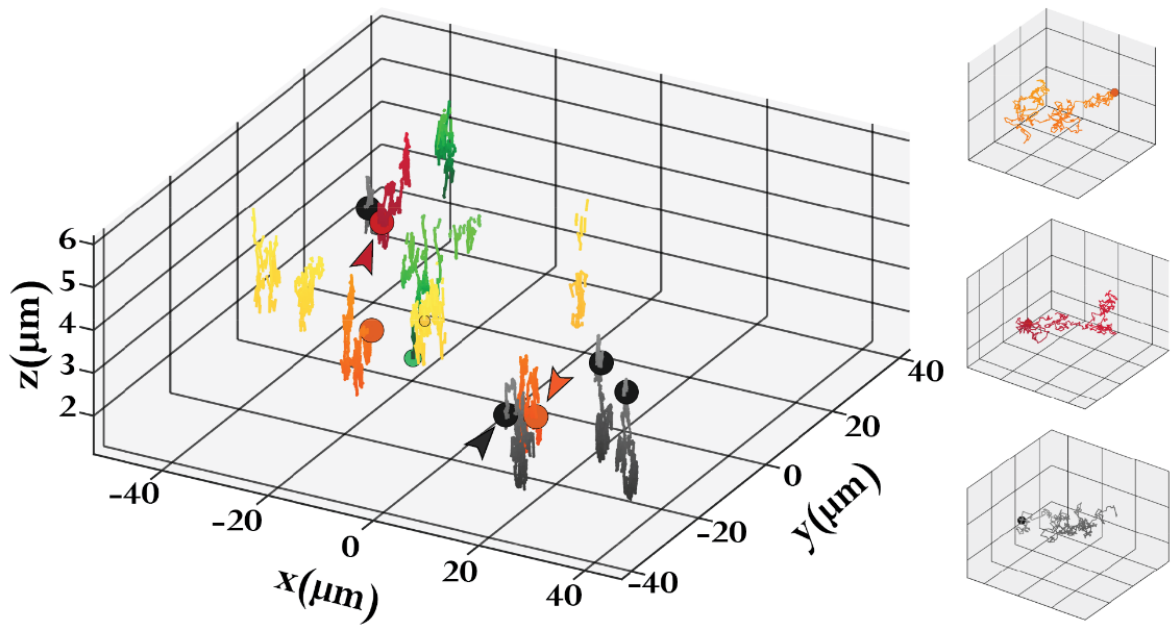

**Fig. 3**: 3D rendered trajectories from a time-lapse acquisition (see Visualization 1). Color and axial position are represented by the trajectory color (brighter shades correspond to higher axial values). The final position of beads that remained inside the FOV at the end of the acquisition are expressed as a sphere at the final measured position. To the right, the trajectories marked by arrowheads, shown with aspect ratio of 1:1:1; Gridlines of the zoomed-in trajectories are 2 μm apart.

### 3.2 Live-cell imaging

To demonstrate the applicability of the method for multicolor imaging of a live biological system, we performed tracking of fluorescently tagged DNA loci in *Saccharomyces Cerevisiae (SC)* yeast cells. Two loci on both sides of the GAL locus were tagged[10] using the Fluorescent Repressor Operator System (FROS) LacO-LacI-eGFP[37] (green) and TetO-tetR-mCherry[38] (red) (Fig. 4b). Their 3D motion was tracked and their trajectories were extracted (Fig 4.c).

Two major challenges that arise in imaging of live cells, compared to the case of freely diffusing beads are cellular background and strong photobleaching. These issues manifest in decreased overall precision: The first requires a more challenging fitting scheme with an increased number of parameters, using a weaker model of an additive 2D Gaussian function (see "Localization" section). The latter reduces detected photon counts for longer cumulative laser exposure duration. In order to assess the localization precision of intra-cell imaging, we imaged fixed cells under similar optical conditions, i.e. illumination and exposure times, to the live cell imaging. The standard deviation of the localizations of each locus was taken as the experimental localization precision. The estimated 3D precision values for {LacO,TetO} were {59.1,25.0}±{20.7,10.1} nm, taken from 16 cells. The mean loci distance measured from tracking 35 live cells (attained from 6 time-lapses) was 282±101 nm, in agreement with previous measurements of this biological system[10].

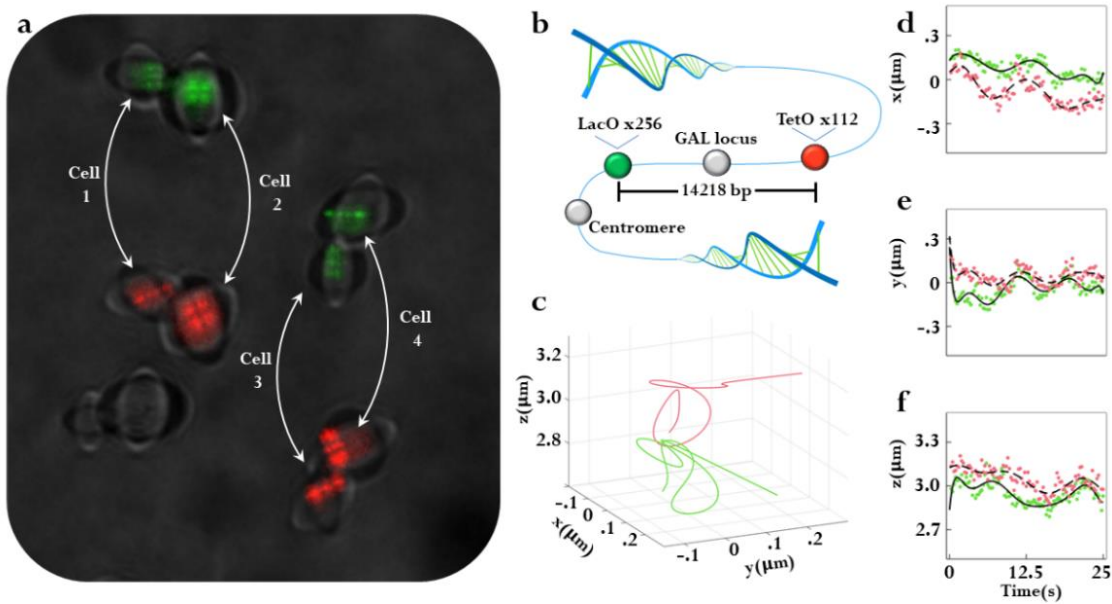

**Fig. 4**: Multicolor *in-vivo* imaging of *S. Saccharomyces* DNA loci. **a)** Cropped FOV of an image out of a time-lapse acquisition of live cells (see Visualization 2.). The PSFs of the fluorescent loci are falsely colored depicting emission color. PSF pairs of the same DNA strand are marked by white connecting arrows. The PSFs are overlaid with a darkened bright-light image of the cells, as seen captured through the optical system. **b)** Labelling scheme of the yeast DNA. **c)** Representative measured trajectories of the fluorescent loci. **d-f)** Scatter plots of x,y,z localizations from the data used for c), green and red represent LacO and TetO respectively. Black lines show polynomial fits of the trajectory shown in 4.c. Continuous line corresponds to LacO, and dashed line to TetO. The origin of the lateral coordinates (x,y) in c-f was chosen arbitrarily for display. The axial coordinates are with respect to the coverslip surface.

In the cell imaging experiments, the signal was very well separated between the blue and orange channels, as indicated by the variability in measured loci trajectories, and also verified by imaging different strains of *S. Saccharomyces,* containing a single type of fluorescent protein (EGFP or mCherry) each. These measurements have shown that while some background is apparent in both channels, induced by the 488 nm laser excitation, there is no trace of localized signal from any loci in the other channel.

## 4. Discussion

Here, we have proposed an approach that overcomes the challenges of fast, multicolor, SPT by eliminating the tradeoff between acquiring spectral information, 3D spatial resolution, temporal resolution, and FOV size. As a proof of concept, we have demonstrated simultaneous, five-color tracking, at camera-rate limited speeds (~50 Hz), with tens of nanometers precision, of >10 objects over a large FOV of 110×70×4 µm$^3$. We applied our technique to characterize the 3D motions of fluorescent nanoparticles *in vitro* and in biological samples, specifically, 3D tracking of fluorescently tagged DNA in live cells. The key advantage of our approach is that it harnesses the additional degree of freedom afforded by PSFE, namely uniquely encoding the color and position into the shape formed on the image, while maintaining a large field of view on a detector. This approach also alleviates the need for synchronized triggering of multiple detectors. Differing from our earlier implementation of single FOV, multicolor PSFE[29], the multi-path design benefits from simpler DOE designs that are significantly more photon efficient and are designed to utilize various parts of the visible-NIR spectrum.

The multiplexed PSFE approach has additional benefits as well, as the PSF in each path can be individually optimized to meet specified requirements that may differ between the channels. For example, PSFs optimized for different axial ranges can be used for complicated biological systems, or extended depth-of-field (EDOF) PSF can be used in a single channel to laterally monitor the position of a cell membrane while tracking single particle inside.

One limitation of our approach is in capturing the positions of emitters that overlap both spatially and spectrally; however, narrowing the individual transmitted spectral windows by filters or changing the fluorescent label can circumvent this issue. For sparse emitters, broad emission spectra can be advantageous, as ratiometric spectral analysis can be used to increase the number of distinguishable emitter types beyond the number of captured channels.

Future improvements to the optical setup and analysis pipeline may enable further increase in the allowable density of emitters. For example, in the live-cell imaging experiments, where there is a known number of emitters in a cell occupying an approximately known area, the configuration could be optimized to match the density of emitters and the spacing between channels as to prevent PSF overlap of adjacent cells. Computationally, employing multi-PSF fitting can further expand the allowable densities. This algorithmic approach would also improve analysis of unconstrained diffusers, as in the current framework occasional PSF overlap hampers the localization process, and results in shorter trajectories for subsequent analysis. For high density, more sophisticated analysis methods can be developed, e.g. using machine learning and possibly exploiting temporal correlations[39,40].

In conclusion, multiplexed PSFE provides a modular and photon efficient solution for multicolor 3D localization with a large FOV. By simultaneously capturing the spectral information on a single camera, we anticipate this technique to be applicable to speed-necessitating experiments in challenging samples including *in vivo* imaging of fluorescent labels, and colocalization of objects in flow[41].

### Acknowledgements

We would like to thank Elisa Dultz and Karsten Weis for sending us the cells used for the live-cell imaging. N.O., Y.S.E., B.F. and R.O. have received funding from the European Union's Horizon 2020 research and innovation program under grant agreement No. 802567 -ERC- Five-Dimensional Localization Microscopy for Sub-Cellular Dynamics. L.E.W. and Y.S. are supported by the Zuckerman Foundation; N.O. is also supported by the Israel Innovation Authority.